\documentclass{iaus}
\usepackage{graphics}

  \checkfont{eurm10}
  \iffontfound
    \IfFileExists{upmath.sty}
      {\typeout{^^JFound AMS Euler Roman fonts on the system,
                   using the 'upmath' package.^^J}%
       \usepackage{upmath}}
      {\typeout{^^JFound AMS Euler Roman fonts on the system, but you
                   dont seem to have the}%
       \typeout{'upmath' package installed. iaus.cls can take advantage
                 of these fonts,^^Jif you use 'upmath' package.^^J}%
      }
  \else
  \fi

  \checkfont{msam10}
  \iffontfound
    \IfFileExists{amssymb.sty}
      {\typeout{^^JFound AMS Symbol fonts on the system, using the
                'amssymb' package.^^J}%
       \usepackage{amssymb}%

      }{}
  \fi

  \IfFileExists{amsbsy.sty}
    {\typeout{^^JFound the 'amsbsy' package on the system, using it.^^J}%
     \usepackage{amsbsy}}
    {}

\newsavebox{\astrutbox}
\sbox{\astrutbox}{\rule[-5pt]{0pt}{20pt}}

\newcommand\etal{\mbox{\textit{et al.}}}

\title[The Interplay among Black Holes, Stars and ISM in Galactic 
       Nuclei]{Comparing AGN\\
broad-- and narrow--line regions}

\author[N. Bennert {\it et al.\/}]%
{Nicola Bennert$^1$, Heino Falcke$^2$, Yuri Shchekinov$^3$, Andrew S. Wilson$^4$}
\affiliation{$^1$AIRUB,
Universit\"{a}tsstra{\ss}e 150, 44780 Bochum, Germany\\$^2$ASTRON, P.O. Box 2,                       
7990 AA Dwingeloo, The Netherlands\\ $^3$Rostov 
State University, Department of Physics, 344090 Rostov on Don,
Russia\\ $^4$Astronomy Department, University of Maryland, College Park, MD
20742--2421, USA}

\pubyear{2004}
\volume{222}
\pagerange{1--8}
\date{?? and in revised form ??}
\jname{The Interplay among Black Holes, Stars and ISM \\in Galactic Nuclei}
\editors{Th. Storchi Bergmann, L.C. Ho \& H.R. Schmitt, eds.}
\begin{document}
\maketitle

\abstract{We compare recent HST
observations of Seyfert and quasar NLRs and find
that type--2 AGNs follow a relation consistent with that
expected for a distribution of gas ionized by a central source 
$R_{\rm NLR,2} \propto L^{0.32 \pm 0.05}$, while type--1 objects
are fit with a steeper slope of $0.55 \pm 0.05$.
The latter is comparable to the scaling found for 
the BLR size with continuum luminosity (slope: 0.5--0.7).
Therefore, we investigate what we can learn about the BLR size
if the NLR size is only determined by the AGN luminosity.
We  find that 
NLR and BLR size are related linearly following $R_{\rm BLR} \propto R_{\rm
NLR,1}^{0.88 \pm 0.1}$. This relation can be used to estimate
BH masses.}

\firstsection 
\section{Summary}
Sizes and morphologies of broad-- and narrow--line regions (BLRs, NLRs) 
provide an ideal probe of the distribution of dust and gas
in the central parts of AGNs.
However, until today, even very basic issues
remain open: What determines the size and structure of these
emission--line regions? Do they grow with luminosity, and if so, how?
For the NLR, the latter was investigated for the first time by 
\cite[Bennert \etal\ (2002)]{ben02}
for a sample of seven radio--quiet PG quasars
observed in the [OIII]\,$\lambda$5007 line with HST. We found 
a NLR size--luminosity relation
$R_{\rm NLR} \propto L_{\rm [OIII]}^{0.5}$
when including a sample
of seven Seyfert--2 galaxies (\cite[Falcke \etal\ 1998]{fal98}).
This result is remarkable if it implies that it is not the 
Str\"omgren radius that limits the NLR but
an apparent threshold in ionizing flux, which can be
expressed by a constant product of ionization parameter
and density at the rim of the NLR. 
It is not clear whether this new relation will
hold at all luminosities and redshifts. If, for example, the
luminosity--scaled size of the NLR becomes larger than the size of the
galaxy, the relation may flatten as the emission lines fade out and
disappear (\cite[Netzer 2004]{net04}). This may be evident already in some spectral line studies
of Two Degree Field quasars (\cite[Croom \etal\ 2002]{cro02}). 

More recently, \cite[Schmitt \etal\ (2003)]{sch03} studied the 
extended [OIII] emission in a 
sample of 60 Seyfert galaxies 
and find that the NLR size--luminosity relation
follows a simple Str\"omgren law
$R_{\rm NLR} \propto L_{\rm [OIII]}^{0.33}$.
What causes the apparent different slopes for our quasar--dominated
sample (0.5) and their Seyfert sample (0.33): simple statistical
uncertainty or selection effects? Are the NLRs of Seyferts and
quasars intrinsically different? Or is there
a difference in the NLR size--luminosity relation between type--1
and type--2 objects? Puzzled by these questions, we compared the two samples and
calculated a fit to all type--1 and type--2 AGNs separately:
While type--1 objects follow the relation 
$R_{\rm NLR,1}
\propto L^{0.55 \pm 0.05}$, type 2s can be fit by
approximately a Str\"omgren--law $R_{\rm NLR,2}
\propto L^{0.32 \pm 0.05}$ (Fig. 1, {\it left}).
This result explains the different slopes found by the two
groups: In the \cite[Bennert \etal\ (2002)]{ben02} sample,
basically type--1 quasars define the slope,
whereas the slope of the \cite[Schmitt \etal\ (2003)]{sch03} sample is dominated by
Seyfert 2s which outnumber the Seyfert 1s by a factor of 1.4. 

While the NLR size can be easily measured from images,
the BLR is too compact to derive its size directly.
Instead, observations of correlated variations in the line and continuum
emission are used to determine the BLR size
indirectly by the reverberation mapping techniques
(e.g. \cite[Wandel \etal\ 1999]{wan99}; \cite[Kaspi \etal\ 2000]{kas00}).
Investigations of the BLR size--luminosity relationship
lead to controversy about the slope:
While \cite[Kaspi \etal\ (2000)]{kas00} report a relation of $R_{\rm BLR} \propto L^{0.7}$,
\cite[McLure \& Jarvis (2002)]{mcl02} find $R_{\rm BLR} \propto L^{0.5}$.
If the NLR size is only determined by the AGN luminosity,
it is of interest what we can learn about 
the BLR size and BH mass.
Therefore, we compared all measured BLR sizes with their
corresponding NLR sizes. 
Unfortunately, the overlap is rather poor and leaves
us with 11 objects.
Applying a weighted linear least--squares fit, we
find that
the sizes of NLR and BLR are 
proportional: $R_{\rm BLR} \propto R_{\rm NLR,1}^{0.88 \pm 0.1}$ (Fig. 1, {\it right}).

\begin{figure}
\centering
\resizebox{6.5cm}{!}{\includegraphics{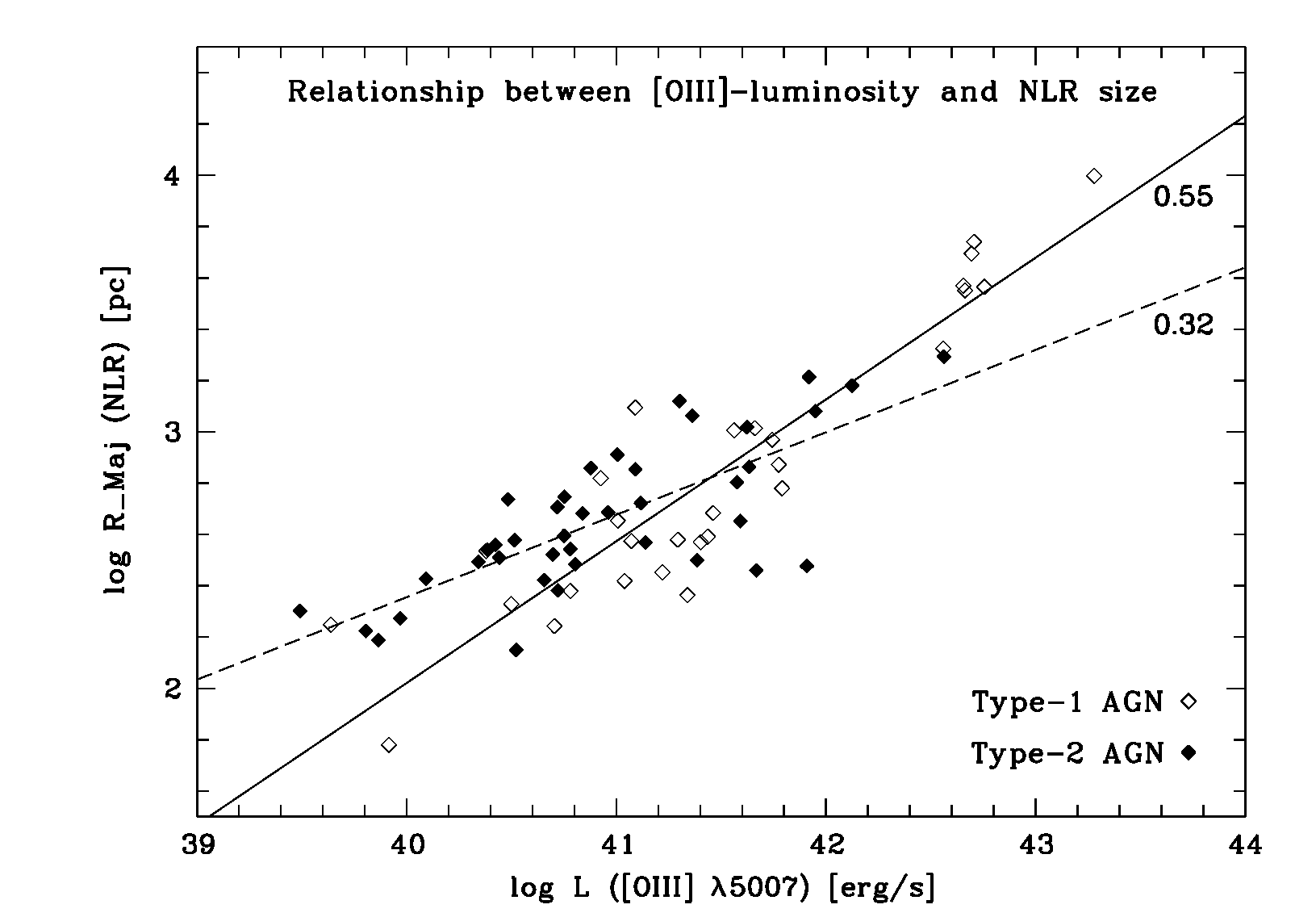} }
\resizebox{6.5cm}{!}{\includegraphics{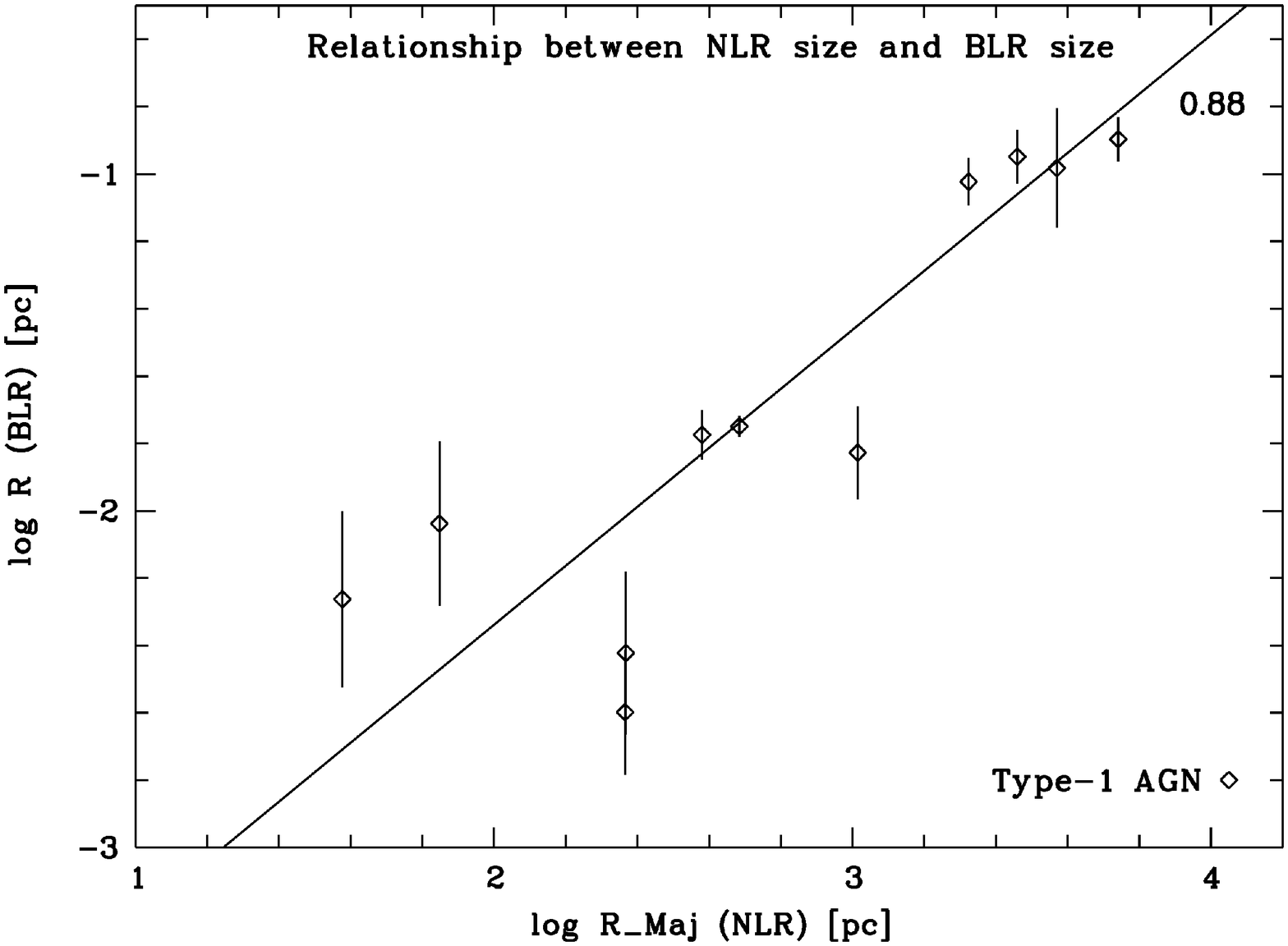} }
\caption[]{{\it Left:} Radius of the NLR
versus the emission--line luminosity in [OIII]
on logarithmic scales.
The solid line represents the fit to type--1 AGNs
($R_{\rm Maj}$ (NLR,1) $\propto L_{\rm [OIII]}^{0.55 \pm 0.05}$),
the dashed line the fit to type--2 AGNs
($R_{\rm Maj}$ (NLR,2) $\propto L_{\rm [OIII]}^{0.32 \pm 0.05}$).
{\it Right:} Distribution of NLR and BLR radii on
logarithmic scales. The error bars indicate the uncertainty in
defining the BLR radius.
The solid line represents the fit corresponding
to $R$ (BLR) $\propto R_{\rm Maj}$ (NLR,1)$^{0.88 \pm 0.1}$.}
\end{figure}

Since deriving BLR sizes via reverberation mapping is very time consuming,
relationships with luminosity or NLR size
can be extremely useful for estimating BH masses. 
Given the NLR size (either directly measured or derived from the 
[OIII] luminosity),
the BLR size can be estimated, which, assuming Keplerian motions, yields $M_{\rm BH}$:
\begin{eqnarray*}
M_{\rm BH} = (10^{5.22} M_{\odot}) v_{3000}^2 R_{\rm NLR,1}^{0.88}
\hspace*{0.5cm} {\rm or} \hspace*{0.5cm}
M_{\rm BH} = (10^{8.83} M_{\odot}) v_{3000}^2 L_{44, \rm [OIII]}^{0.48}
\end{eqnarray*}
(where $v_{3000}$ = FWHM (H$\beta$)/3000
km\,s$^{-1}$, $R_{\rm NLR,1}$ in parsec, and
$L_{44, \rm [OIII]}$ =  [OIII]
luminosity/$10^{44}$ erg\,s$^{-1}$).
The latter equation 
is very similar to the correlation of $M_{\rm BH}$ and continuum
luminosity at 5100 \AA~(\cite[Shields \etal\ 2003]{shi03}).

The observed slopes of the NLR size--luminosity 
and the BLR--NLR--size relation
can be explained by a simple model based on
the unified scheme and the receding torus model, 
taking into account the NLR size dependency on viewing angle
and luminosity (\cite[Bennert \etal\ 2004]{ben04}).

\acknowledgements
In memoriam Prof.~Hartmut Schulz, deceased in August 2003.
N.B. remembers him gratefully for having been her PhD advisor and
``Doktorvater'' in the truest sense of the word.
The astronomical community has lost a wonderful colleague and a truly
independent mind.


\begin{thebibliography}{}

\bibitem[Bennert \etal\(2002)]{ben02}
{Bennert, N., Falcke, H., Schulz, H.,
Wilson, A. S., \& Wills, B. J.} 2002, \textit{ApJL}, \textbf{574}, 105

\bibitem[Bennert \etal\(2004)]{ben04}
{Bennert, N., Shchekinov, Y., Falcke, H., 
\& Wilson, A. S.} 2004, in preparation 

\bibitem[Croom et al.(2002)]{cro02}{Croom, S. M., Rhook, K., 
Corbett, E. A. \etal\ } 2002, \textit{MNRAS}, \textbf{337}, 275

\bibitem[Falcke \etal\(1998)]{fal98} 
{Falcke, H., Wilson, A. S., 
\& Simpson, C.} 1998, \textit{ApJ}, \textbf{502}, 199

\bibitem[Kaspi \etal\(2000)]{kas00}{Kaspi, S., Smith, P. S., Netzer, H.
\etal\ } 2000, \textit{ApJ}, \textbf{533}, 631

\bibitem[McLure \& Jarvis(2002)]{mcl02}{McLure, R. J. \& Jarvis, M. J.} 2002,
\textit{MNRAS}, \textbf{337}, 109

\bibitem[Netzer (2004)]{net04}{Netzer, H.} 2004, this volume

\bibitem[Schmitt \etal\(2003)]{sch03}{Schmitt, H. R., Donley, J. L., 
Antonucci, R. R. J. \etal\ } 2003, \textit{ApJS}, \textbf{597}, 768

\bibitem[Shields \etal\(2003)]{shi03}{Shields, G. A., Gebhardt, K., Salviander,
  S. \etal\ } 2003, \textit{ApJ}, \textbf{583}, 124

\bibitem[Wandel \etal\(1999)]{wan99}{Wandel, A., Peterson, B. M., \& Malkan, 
M. A.} 1999, \textit{ApJ}, \textbf{526}, 579
\end{thebibliography}
\end{document}